# MULTIDIMENSIONAL ANALYSIS OF MONTHLY STOCK MARKET RETURNS

**Osman GULSEVEN**[*]

**Abstract**

*This study examines the monthly returns in Turkish and American stock market indices to investigate whether these markets experience abnormal returns during some months of the calendar year. The data used in this research includes 212 observations between January 1996 and August 2014. I apply statistical summary analysis, decomposition technique, dummy variable estimation, and binary logistic regression to check for the monthly market anomalies. The multidimensional methods used in this article suggest weak evidence against the efficient market hypothesis on monthly returns. While some months tend to show abnormal returns, there is no absolute unanimity in the applied approaches. Nevertheless, there is a strikingly negative May effect on the Turkish stocks following a positive return in April. Stocks tend to be bullish in December in both markets, yet we do not observe anya significant January effect is not observed.*

## 1. INTRODUCTION

The stock markets are among the most efficient ones where thousands, if not millions, of buyers and sellers act almost instantly to any new information. As such, one would expect that any abnormal opportunity to disappear as soon as it is discovered by investors, hedgers, or arbitrageurs. There is a tendency for the stock market indices to grow over time, but there is an almost absolute uncertainty in guessing direction of the market index tomorrow. Yet, there is a growing amount of interest on explaining and forecasting the future movements of the stock markets. Seasonal anomalies are among the several approaches applied by investors who try to achieve better than average returns. Day of the week effect, turn of the month effect, Christmas holiday effect, positive December-January effect, negative May effect are among the most widely researched subjects. In this article, I look for monthly seasonal effects.

The existences of monthly seasonal effects are tested in many different markets. January effect is one of the most investigated seasonal behaviours (Thaler, 1987). Wachtel (1942) was among the first researchers who investigated monthly seasonalities. His research suggested that markets have a tendency to be bullish in December and January months. Haugen and Jorion (1996), Mehdian and Perry (2002) also reported results that confirm the existence of these seasonalities. Gultekin and Gultekin (1983) suggested that the monthly seasonal effects are not limited to US markets only. The authors find disproportionately large returns in most countries during the turn of the tax year period. Contrary to this result, Fountas and Segredakis (2002) report very little evidence in favor of the January effect and the tax-loss selling hypothesis. Similar research has been done in Japan (Kato and Schalleim, 1985) and Kuwait (Al-Saad and Moosa, 2005). Using 25 years of recent data in Bangladesh market, Ahsan and Sarkar (2013) do not find a significant January effect, but they discover a significantly positive return in June. So, the evidence for and against the seasonal factors is mixed (Agrawal and Tandon, 1994).

---

[*] Faculty of Economic and Administrative Sciences, Middle East Technical University, Çankaya Ankara, Turkey, e-mail: *gulseven@metu.edu.tr*.

Another widely disputed calendar anomaly is the negative May affect. "Sell in May and go away" is a common jargon in the finance industry. This strategy, popularized by O'Higgins and Downes (2000) is still an ambiguous one. Bouman and Jacobsen (2002) found strong evidence for May effect in 36 of 37 countries in their sample data. Dzhabarov and Ziemba (2010) suggest that the Sell-in-May-and-go-away phenomenon tends to be stronger in small-cap stocks.

The calendar effects are also studied in the Turkish markets as well. Dicle and Hassan (2007) suggest that Mondays have negative returns whereas Thursday and Friday tend to be positive. The turn-of-the-month effect has been studied by Oguzsoy and Guven (2006). The authors find evidence in favour of this effect if the turn of the month is defined as the day before and three days after the first day of the month. Past research on Turkish data also indicates significant monthly anomalies, which are particularly evident in January (Eken and Üner, 2010). Bildik (2004) suggests that the monthly calendar anomalies are not only present in the stock return but also in trading volume.

These empirical studies utilize the past data to test their model. One of the problems in some of these studies is the data mining issue. It is a common practice to first check the data and then try to develop the model that fits best with the existing data. Here, I apply a completely unbiased approach using four different methods based on statistical, econometric, and probabilistic analyses. This article's objective is to determine whether it is possible to guess the future movement of the stock market on a monthly basis. The organization of the article is as follows. First, I briefly explain the sources of the data and how the data is transformed into different forms to fit the techniques used in this article. Next, the applied models are discussed in the methods section. After explaining the results for each technique, I summarize the results in the summary section. Finally, the discussion section offers a brief behavioural analysis of the stock markets.

## 2. DATA

Data is derived from publicly available databases. For the US market, I utilized the S&P 500 ETF data which tracks the S&P 500 index. This index tracks the performance of broad domestic stocks and is widely used in market analysis. S&P 500 data is downloaded from Yahoo Finance website. The data is adjusted for dividends and splits. It covers the period from January 1996 to August 2014.

For the Turkish market, I utilized the BIST 30 national index which tracks the performance of domestic stocks. These stocks are selected based on their market cap in the market. The data is downloaded from Borsa Istanbul website. In the past, this index was named as IMKB 30 index. With the recent changes in the status of the stock markets, it is renamed as BIST 30 index. This index is defined in terms of Turkish Lira, US Dollar, and also European Euro. I choose the USD-based index since the Turkish stock market is dominated by foreign investors. The USD-based BIST 30 index also provides harmony with the S&P 500 data as it is also denoted in USD. Moreover, the USD-denoted index fits better for the purpose of our analysis as there are several factors which affect the value of Turkish Lira against USD. The TL-based index is subject to extreme volatilities such as currency devaluations that are beyond the scope of this article. As it is the case that between 60 to 70 percent of the Turkish stocks is owned by investors of non-Turkish origin using USD-based index is a logical step. Of course, it is possible to utilize the TL-based data, but then factors that affect currency valuations would also need to be considered. As I am looking for only time-based factors affecting stock markets, I used the USD-based index.

Before moving a step further, one needs to decide whether it is feasible to use the data in nominal form or whether there some modifications are needed in the data. This is particularly important in econometric analysis where stationarity is a pre-condition for robust econometric results.

First, I checked for the amount of autocorrelation in the monthly data using the calculate autocorrelation function in the Minitab software. The autocorrelation and partial autocorrelation functions for the nominal indices suggested a significant autocorrelation over time, where the nominal values are highly autocorrelated with each other.

The autocorrelation and partial autocorrelation functions in Figure 1 suggested an AR(1) model, which can be defined as $X_t = \delta + \varphi_1 X_{t-1} + w_t$. In the current form, one can utilize this data only if the model allows for autocorrelation. When applying decomposition analysis these issues are automatically solved thanks to the trend function that captures this behavior. Therefore, nominal data is used in decomposition analysis as it allows for existence of trend and accommodates autocorrelation in the data.

While decomposition analysis allows for such autocorrelated series, one needs to have stationary data in order to apply dummy variable regression and binary logistic models. There are several ways to achieve stationarity. I apply the simplest approach by differencing the data and transforming the nominal values into percentage returns. This way the autocorrelation issues automatically disappeared.

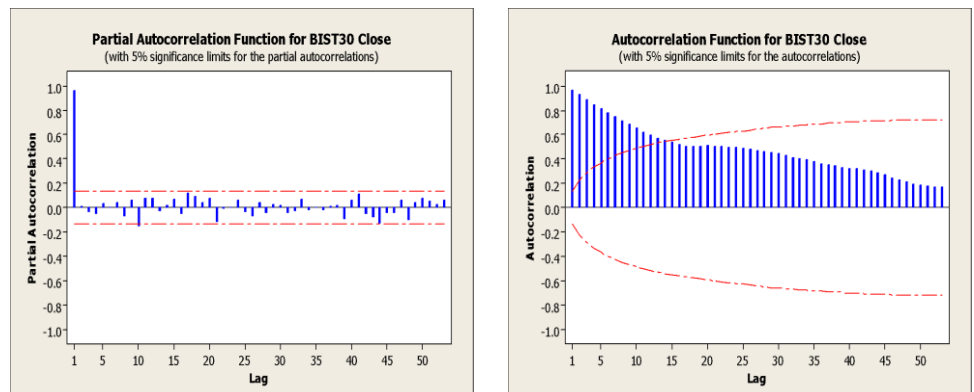

**Figure no. 1 – Autocorrelation and partial autocorrelation functions for nominal data values**

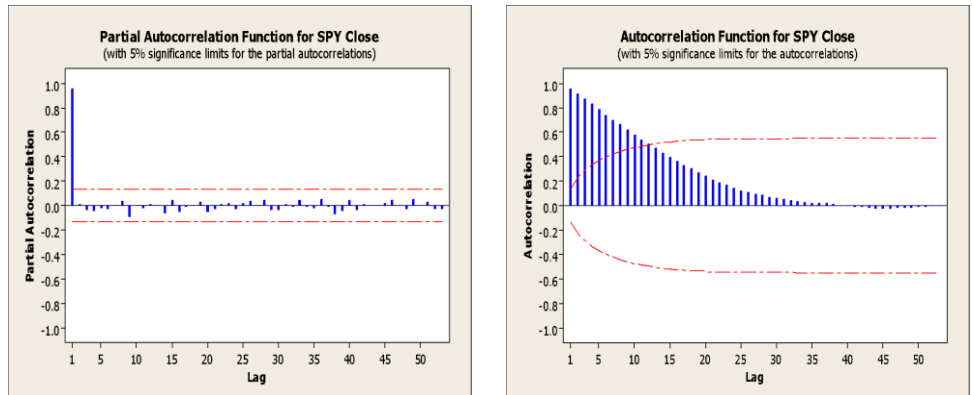

**Figure no. 1 – Continued**

Using the "month" formula within Excel 2007, the data is categorized based on which month it is recorded. For both markets the nominal monthly performance is calculated as the difference between the closing price of the last trading day of the previous month and the last trading day of the current month. The percentage return is calculated by dividing this difference with the index value in the last trading day of the previous month.

$$\% \text{ Return} = 100 \times (\text{Index Value}_t - \text{Index Value}_{t-1}) / \text{Index Value}_{t-1}$$

In the above form, the explained variable is the monthly percentage return. As can be seen in Figure 2, the data in the monthly percentage return form is almost completely random. Moreover, both the autocorrelation and partial autocorrelation functions show stationary behavior.

The monthly percentage returns are utilized to calculate statistical summary variables by each month. They are also used in the dummy variable regression and binary logistic model. The nominal index values are used only in decomposition analysis which allows for such autocorrelation in the data.

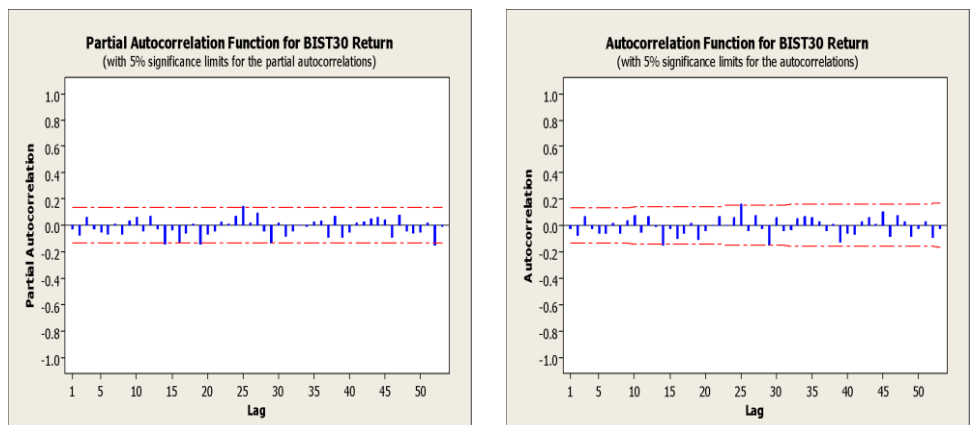

**Figure no. 2 – Autocorrelation and partial autocorrelation functions for monthly percentage returns**

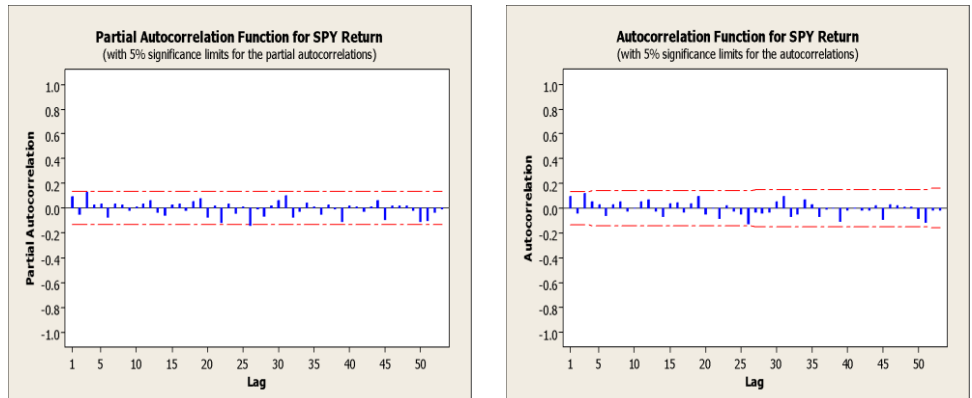

**Figure no. 2 – Continued**

12 binary dummy variables are created for each month. For example, the dummy on January equals one if the recording month is January. Otherwise, the January dummy would be 0. Those dummy variables are used in the dummy variable regression. They are also used in the binary logistic model, where I estimate the possibility of a positive return based on monthly dummies. When estimating the binary choice model, the monthly returns are also converted into binary variables. If the return in a specific month is positive, then the value on this binary will be equal to 1, if not, it will be equal to 0. Here is a basic statistical summary of the data used in estimation:

**Table no. 1 – Statistical summary of BIST 30 and S&P 500 index monthly returns**

|  | Count | Minimum | Maximum | Mean | Median | Std. Dev. | 95% LCL | 95% UCL |
|---|---|---|---|---|---|---|---|---|
| BIST | 212 | -40.16 | 72.00 | 1.89 | 1.87 | 15.30 | -0.17 | 3.96 |
| SPY | 212 | -16.52 | 10.92 | 0.72 | 1.27 | 4.52 | 0.11 | 1.33 |

The above summary suggests interesting results. The average monthly return in BIST 30 index is 1.89%, whereas this return falls to 0.72% for S&P 500. However, the 95% confidence interval for S&P 500 does not include 0, whereas 0 is included in this interval for BIST 30. Based on the data one can expect a statistically significant positive return for S&P 500, but we reject statistically significant positive returns in BIST 30. This is an interesting result as the Turkish stocks performed significantly better than the American stocks during the research period. However, there is a logical explanation for this observation. The Turkish data shows a higher expected return whereas S&P 500 data has substantially lower volatility. Thus, this observation in data is in line with the generally accepted financial principles, where more return comes with higher risk..

## 3. METHODS

### 3.1. Statistical Summary Analysis

In terms of statistical analysis, I used the "calculate basic statistics" option available in most statistical analysis software. However, separate statistics are calculated for each month. This way, the return for each month is denoted as a separate variable. The mean, standard deviation, the t-values and 95% confidence intervals are calculated separately for each month. The 95% confidence intervals for mean are calculated using mean plus/minus margin of error formula.

### 3.2. Decomposition Analysis

The decomposition analysis is one of most practical approaches when it comes to forecasting data that shows seasonal behaviour. This approach is applied in mostly macroeconomic time series analysis (Nelson and Plosser, 1982; Gooijer and Hyndman, 2006), but it also has several applications in industry (Grubb and Mason, 2001; Segura and Vercher, 2001). The decomposition analysis is also used in biometric models (West, 1997).

In the decomposition analysis the data is theoretically divided into four components: Trend, Seasonality, Cycle, and Irregular components. The trend component captures the upward or downward trends, whereas the seasonal component tries to capture pre-defined seasonal effects in the data. Since this paper aims to capture abnormal monthly returns, the seasons are defined in terms of each month. The stock market data follows somewhat a similar cycle to that of other macroeconomic variables. However, as it would be impossible to define the cyclical periods, this component is ignored in software application.

There are two ways to decompose the data. In the first approach, the data is decomposed using additive method. Here the model can be defined as Index = Trend + Seasonality + Irregular components. This approach assumes an additive trend and seasonal component which is not logical in the case of exponentially growing stock market indices. Therefore, I utilized the second approach where Index = Trend x Seasonality x Irregular components. Known as the multiplicative decomposition technique, this functional form has a better foundation in the case of exponentially growing variables such as nominal market indices.

### 3.3. Dummy Variable Regression

Unlike the decomposition analysis, in the dummy variable regression model, I utilized the monthly return data as it does not show any sign of autocorrelation or partial autocorrelation. The model assumes a stationary data and I achieved this stationarity by transforming the data into monthly percentage returns. The dummy variable regression coefficient for each month has a value of 1 or 0 depending on whether the data belongs to that month or not. In order to eliminate perfect multicollinearity, I removed the dummy for the December month.

First, all monthly dummy variables excluding the December dummy are included in the model. The full model is shown as below:

$$\text{Return}_t = \beta_0 + \beta_1 D_1 + \beta_2 D_2 + \beta_3 D_3 + \beta_4 D_4 + \beta_5 D_5 + \beta_6 D_6 + \beta_7 D_7 + \\ + \beta_8 D_8 + \beta_9 D_9 + \beta_{10} D_{10} + \beta_{11} D_{11}$$

The full model includes all variables regardless of their actual effect on the return. Since it also includes possibly irrelevant variables, it would be better to include only the coefficients that are most relevant to the returns. One commonly used method is to apply stepwise regression, which has a wide range of applications (Shanableh and Assaleh, 2010). In this method the algorithm first chooses the variable that has the highest correlation with the return and then looks for the next variable that has a high correlation with the explained variable but low correlation with the previously included explanatory variables. The algorithm goes on until there is no significant variable left to add to the model.

The results of the stepwise variable selection suggested the following dummy variable regressions for the BIST 30 and S&P 500 returns. By construction, all of the selected parameter estimates are significant with a 95% confidence.

$$\text{BIST 30 Return}_t = \beta_0 + \beta_1 \times \text{April} + \beta_2 \times \text{May} + \beta_3 \times \text{August} + \beta_4 \times \text{December}$$

For S&P 500 we have a much more basic outcome:
$$\text{S\&P 500 Return}_t = \beta_0 + \beta_1 \times \text{April}$$

### 3.4. Binary Logistic Model

Perhaps the most intuitive method to test for monthly anomalies is the binary logistic model. This model has wide applications in identifying possible factors effective in the occurrence of an event (Sze *et al.*, 2014). While the details of the model might seem complicated, the interpretation of the results is pretty simple. In the binary logistic model we try to calculate the possibility of earning a positive return for each month. First, we classify each event as a success (1) if we earn a positive return and a non-success (0) if we earn a negative return. Next, we look for possible factors that might increase or decrease the chances of earning a positive return.

I defined a new event Return-Positive variable which is equal to 1 for months with positive returns and 0 for months with negative returns. This categorical variable is estimated using a probabilistic approach, where the probability distribution function is defined as follows:

$$F(\text{event}) = \exp(\text{factor function})/[1 + \exp(\text{factor function})] = 1 / [1 + \exp(-\text{factor function})]$$

In this article, I am looking for the calendar of the month effect, so I included the monthly dummies in the estimation. The factor function is defined as:

$$\text{Factor function} = \beta_0 + \beta_1 D_1 + \beta_2 D_2 + \beta_3 D_3 + \beta_4 D_4 + \beta_5 D_5 + \beta_6 D_6 + \beta_7 D_7 + \beta_8 D_8 + \beta_9 D_9 + \beta_{10} D_{10} + \beta_{11} D_{11}$$

Similar to the dummy variable estimation, I excluded the dummy variable for the month of December in order to eliminate perfect multicollinearity. In this form, the logistic model provides a convenient probability function for the probabilistic approach as it takes a probabilistic value between 0 and 1.

### 4. RESULTS

#### 4.1. Statistical Summary Analysis Results

The monthly statistical analysis of BIST 30 data suggests that the market shows seasonal behaviour in some of the months. These market anomalies are particularly evident in the April and July months where a positive return is very likely to occur. Interestingly those positive returns are followed by significantly negative returns. May and August are the months where staying in the market is likely to cost money. It is reasonable to observe a negative return in May as this is the period where most Turkish stocks offer dividends, thereby declining the index return. However, it is surprising to see a significantly negative return in August. January and December offer above average returns, but we fail to reject the hypothesis that January and December have neutral returns under 95% confidence level. Only if we reduce the confidence level to 90%, we can claim a significantly positive return for December.

**Table no. 2 – BIST 30 data statistical analysis of monthly returns**

| Month | Mean  | SE (Mean) | 95% LCL | %95 UCL | Tendency | Significance |
|-------|-------|-----------|---------|---------|----------|--------------|
| Jan   | 3.90  | 4.21      | -4.35   | 12.15   | Positive | No           |
| Feb   | -0.92 | 4.05      | -8.86   | 7.03    | Negative | No           |
| Mar   | -0.43 | 3.03      | -6.36   | 5.51    | Negative | No           |
| Apr*  | 8.39  | 3.86      | 0.82    | 15.96   | Positive | Yes          |
| May*  | -5.61 | 2.69      | -10.88  | -0.34   | Negative | Yes          |
| Jun   | -0.79 | 2.31      | -5.32   | 3.73    | Negative | No           |
| Jul*  | 5.62  | 2.69      | 0.34    | 10.89   | Positive | Yes          |
| Aug*  | -5.22 | 2.81      | -10.34  | -0.11   | Negative | Yes          |
| Sep   | 2.14  | 3.37      | -4.46   | 8.74    | Positive | No           |
| Oct   | 4.80  | 3.30      | -1.67   | 11.27   | Positive | No           |
| Nov   | 2.17  | 4.24      | -6.14   | 10.47   | Positive | No           |
| Dec   | 8.91  | 5.15      | -1.19   | 19.01   | Positive | No           |

*\* Significant with a 95% confidence level.*

**Table no. 3 – S&P 500 data statistical analysis of monthly returns**

| Month | Mean  | SE (Mean) | 95% LCL | %95 UCL | Tendency | Significance |
|-------|-------|-----------|---------|---------|----------|--------------|
| Jan   | 0.06  | 0.99      | -1.88   | 2       | Positive | No           |
| Feb   | -0.22 | 1.07      | -2.33   | 1.88    | Negative | No           |
| Mar*  | 1.85  | 0.9       | 0.08    | 3.61    | Positive | Yes          |
| Apr   | 2.33  | 1.01      | 0.35    | 4.32    | Positive | No           |
| May   | 0.39  | 0.93      | -1.42   | 2.2     | Positive | No           |
| Jun   | -0.14 | 0.92      | -1.94   | 1.66    | Negative | No           |
| Jul   | 0.51  | 1.01      | -1.47   | 2.48    | Positive | No           |
| Aug   | -0.83 | 1.14      | -3.08   | 1.41    | Negative | No           |
| Sep   | -0.36 | 1.41      | -3.13   | 2.42    | Negative | No           |
| Oct   | 1.71  | 1.5       | -1.24   | 4.66    | Positive | No           |
| Nov   | 1.65  | 1.07      | -0.45   | 3.74    | Positive | No           |
| Dec*  | 1.72  | 0.71      | 0.33    | 3.11    | Positive | Yes          |

*\* Significant with a 95% confidence level.*

As expected, the S&P 500 data shows more stability than the BIST 30 index. Similar to BIST 30 data, we observe a significant December effect. The December effect is more evident in the S&P 500 data. It is interesting to observe this positive December phenomena although it is a well discussed stock market anomaly. Under efficient market hypothesis, we would expect speculators to enter the market before December so that they can benefit from this phenomenon. As many participants enter the market, the opportunity to earn abnormal returns shall disappear, but this has not been the case in the observed data. We also observe a strong March where the stock index posts significant positive returns. The negative May effect is not observed. August seems to be a negative month, but this is not statistically significant. So, the only market anomalies we observe in the S&P 500 index is the strongly positive March and December effects.

**4.2. Decomposition Analysis Results**

The decomposition analysis allows separating trend and seasonal effects in the nominal data values. Both additive and multiplicative decomposition results are empirically tested. Since the multiplicative decomposition approach suggested a lower Mean Square Error, only the results from multiplicative decomposition analysis is reported in this analysis. It is also more intuitive to use the multiplication decomposition model in an exponentially growing data. The analysis on the nominal indices suggested the following results:

Table no. 4 – BIST 30 and S&P 500 monthly decomposition analysis results*

|      | Jan  | Feb  | Mar  | Apr  | May  | Jun  | Jul  | Aug  | Sep  | Oct  | Nov  | Dec  |
|------|------|------|------|------|------|------|------|------|------|------|------|------|
| BIST | 1.06 | 1.07 | 1.04 | 1.03 | 1.07 | 0.98 | 0.94 | 0.96 | 0.89 | 0.94 | 1    | 0.98 |
| S&P  | 1.02 | 1.01 | 1.00 | 1.01 | 1.02 | 1.01 | 1.01 | 1.00 | 0.98 | 0.97 | 0.98 | 0.99 |

*BIST 30 Multiplicative Decomposition, Trend Equation: BIST 30 = 479.2 + 11.6*t*
*S&P 500 Multiplicative Decomposition, Trend Equation: S&P 500 = 68.99 + 0.346*t*

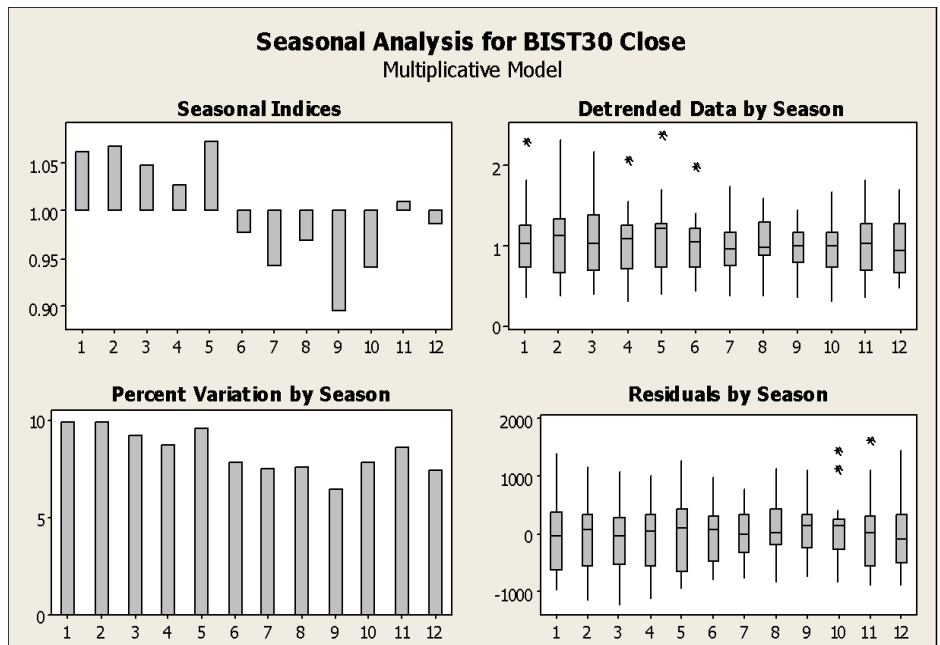

**Figure no. 3 – BIST 30 monthly multiplicative decomposition analysis**

The multiplicative analysis decomposes the data into seasonal and trend components. As expected, the trend component is positive for both BIST 30 and S&P 500 data. Thus, these indices tend to grow over time. We observe similar seasonal components in both indices where the first half of the year performs better than the second half of the year. However, this is likely due to the effect of positive trend component.

The decomposed visualization of data in the above Figure 3 suggests substantially different results from the monthly statistical values. For BIST 30 data, the multiplicative decomposition suggests above average returns until May. Those above average returns disappear as the year end reaches. There is a strikingly negative return on September.

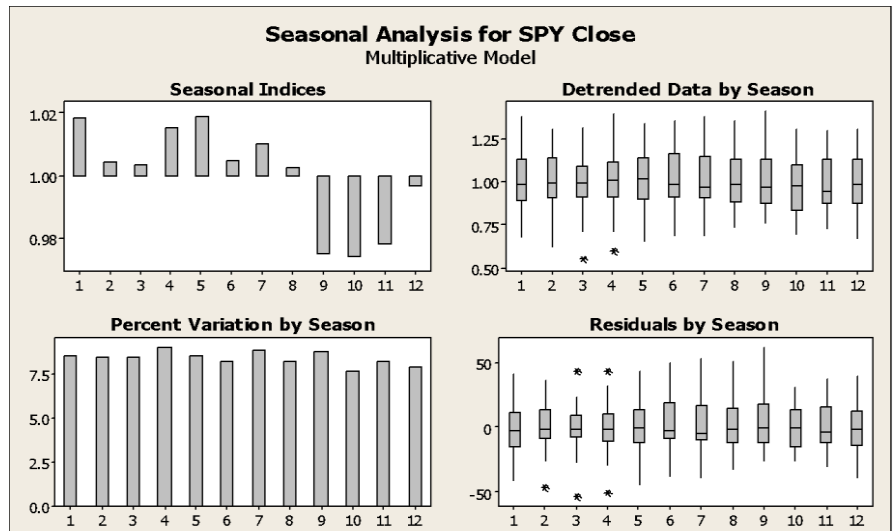

**Figure no. 4 – S&P 500 monthly multiplicative decomposition analysis**

Similar but more stable and less volatile seasonal components are observed for S&P 500 data decomposition. Figure 4 suggests a strong negative index component on September, October, and November. Given that decomposition is one of the well-known methods to analyze seasonal data, the monthly decomposition method does not reveal any visible seasonal behavior in the US stock market. Even the volatility of the data is very similar across different months.

**4.3. Dummy Variable Regression Results**

As discussed in the methods section the dummy variable regression involves creating a linear regression model where the estimators are monthly dummy variables. After creating those dummy variables, I included all the months from January to November in the regression. However, many of the added dummies turned out to be insignificant and thereby irrelevant to the model. Therefore, a stepwise regression technique is utilized to include only the most relevant and thereby significant periods. Both forward and backward selection methods offered the same set of variables to be included in the model.

With an alpha of 0.15, after 4 steps, I obtained the following outcome for the BIST 30 market:

$$\text{BIST 30 Return} = 2.1 + 6.3 \times \text{April} - 7.7 \times \text{May} - 7.3 \times \text{August} + \\ + 6.3 \times \text{December} \ (R\text{-square} = 0.07)$$

All of the parameter estimates are significant at a 95% confidence. The explanatory power of the model is pretty weak as we are using only monthly time dummies to explain the stock market return. Nevertheless, the inclusion of those parameters can give an edge for some investors. The results obtained are mostly in line with the statistical analysis results. April and December offer significantly higher returns, whereas May and August offer significantly lower returns.

Following the same procedure in the analysis of S&P 500 data, I first included all of the monthly dummy variables excluding December. As I obtained insignificant parameter estimates, I used the stepwise regression technique for choosing the best set of explanatory variables. The stepwise regression for S&P 500 data offered a much more basic outcome:

$$\text{S\&P 500 Return} = 0.57 + 1.8 \times \text{April} \ (R\text{-square} = 0.02)$$

The R-square value is almost zero suggesting that seasonal components (if they exist) are a very minor factor in the stock market return. It also suggests that S&P 500 is a relatively more efficient market where the only monthly seasonal component is the April affect. According to the regression above, the stocks offer significantly higher returns in April.

### 4.4. Binary Logistic Model Results

The binary logistic model is a probabilistic model where the explained variable is the probability of the event. Here, I look for whether there are any seasonal factors that can affect the probability of a earning a positive return. The table below shows the logistic estimation results for BIST 30 data.

**Table no. 5 – BIST 30 binary logistic regression results**

| Predictor | Logistic Coefficient | SE Coefficient | Z value | P value | Odds Ratio | 95% LCL | 95% UCL |
|---|---|---|---|---|---|---|---|
| *Constant* | *0.69* | *0.50* | *1.39* | *0.17* | | | |
| Jan | -0.24 | 0.70 | -0.35 | 0.73 | 0.79 | 0.20 | 3.07 |
| Feb | -0.69 | 0.69 | -1.01 | 0.31 | 0.50 | 0.13 | 1.92 |
| Mar | -0.47 | 0.69 | -0.68 | 0.50 | 0.62 | 0.16 | 2.41 |
| Apr | 0.26 | 0.73 | 0.36 | 0.72 | 1.30 | 0.31 | 5.39 |
| May* | -1.39 | 0.71 | -1.96 | 0.05 | 0.25 | 0.06 | 1.00 |
| Jun | -0.92 | 0.69 | -1.33 | 0.18 | 0.40 | 0.10 | 1.54 |
| Jul | 0.00 | 0.71 | 0.00 | 1.00 | 1.00 | 0.25 | 4.00 |
| Aug* | -1.39 | 0.71 | -1.96 | 0.05 | 0.25 | 0.06 | 1.00 |
| Sep | -0.34 | 0.70 | -0.48 | 0.63 | 0.71 | 0.18 | 2.83 |
| Oct | 0.18 | 0.73 | 0.25 | 0.80 | 1.20 | 0.29 | 5.02 |
| Nov | -0.81 | 0.70 | -1.16 | 0.25 | 0.44 | 0.11 | 1.74 |

*Significant with a 95% confidence level.*

The December coefficient is removed in order to eliminate perfect multicollinearity. Therefore, the above results show the probabilistic outcome compared to the reference of December return. In the logistic model, December is obviously a positive performer. Only October and April seem to outperform the reference return. The odds of receiving a positive return in these months are higher than December. However, those positive returns are not statistically significant. The significant abnormal returns happen to be only in the months of May and August. In May and August, the odds of earning a positive return are significantly below than that of December. This result is in line with both statistical analysis and the dummy variable regression I applied previously.

Table no. 6 – S&P 500 binary logistic regression results

| Predictor | Logistic Coefficient | SE Coefficient | Z Value | P Value | Odds Ratio | 95% LCL | 95% UCL |
|---|---|---|---|---|---|---|---|
| *Constant* | *1.25* | *0.57* | *2.21* | *0.03* | | | |
| **Jan** | -1.03 | 0.74 | -1.39 | 0.16 | 0.36 | 0.08 | 1.52 |
| **Feb** | -1.03 | 0.74 | -1.39 | 0.16 | 0.36 | 0.08 | 1.52 |
| **Mar** | -0.30 | 0.77 | -0.38 | 0.70 | 0.74 | 0.16 | 3.38 |
| **Apr** | -0.30 | 0.77 | -0.38 | 0.70 | 0.74 | 0.16 | 3.38 |
| **May** | -1.25 | 0.74 | -1.70 | 0.09 | 0.29 | 0.07 | 1.21 |
| **June** | -1.03 | 0.74 | -1.39 | 0.16 | 0.36 | 0.08 | 1.52 |
| **July*** | -1.48 | 0.74 | -2.00 | 0.05 | 0.23 | 0.05 | 0.97 |
| **Aug** | -1.03 | 0.74 | -1.39 | 0.16 | 0.36 | 0.08 | 1.52 |
| **Sep** | -0.90 | 0.75 | -1.19 | 0.23 | 0.41 | 0.09 | 1.78 |
| **Oct** | -0.65 | 0.76 | -0.85 | 0.40 | 0.52 | 0.12 | 2.33 |
| **Nov** | -0.38 | 0.78 | -0.49 | 0.63 | 0.69 | 0.15 | 3.15 |

*\*Significant with a 95% confidence level.*

Similar to the BIST 30 model, the binary logistic regression on S&P 500 takes the December monthly return as the reference point. The model suggests that the odds ratio for obtaining a positive return is lower in any month compared to the return in December. Thus, December is likely to be an outperformer among other months. The negative coefficient on July return is the highest in absolute magnitude. Given the relatively lower standard errors in the S&P 500 data, this result implies a significantly lower return in the month of July for the US markets.

### 5. SUMMARY

In this article, I applied four different methods to estimate the monthly returns for both BIST 30 and S&P 500 data. Each method is based on different assumptions and as such utilizes different techniques. The results are summarized as below.

Table no. 7 – Summary of results (BIST 30, S&P 500)

| Monthly Return | Logistic Regression | Dummy Regression | Decomposition Analysis | Statistical Results |
|---|---|---|---|---|
| **Jan** | (Pos, Neg) | (None, None) | (Pos, Pos) | (Pos, Pos) |
| **Feb** | (Neg, Neg) | (None, None) | (Pos, Pos) | (Neg, Neg) |
| **Mar** | (Pos, Pos) | (None, None) | (Pos, Pos) | (Neg, Pos*) |
| **Apr** | (Pos, Pos) | (Pos*, Pos*) | (Pos, Pos) | (Pos*, Pos) |

| Monthly Return | Logistic Regression | Dummy Regression | Decomposition Analysis | Statistical Results |
|---|---|---|---|---|
| **May** | (Neg*, Neg) | (Neg*, None) | (Pos, Pos) | (Neg*, Pos) |
| **June** | (Neg, Neg) | (None, None) | (Neg, Pos) | (Neg, Neg) |
| **July** | (Pos, Neg*) | (None, None) | (Neg, Pos) | (Pos*, Pos) |
| **Aug** | (Neg*, Neg) | (Neg*, None) | (Neg, Pos) | (Neg*, Neg) |
| **Sep** | (Pos, Neg) | (None, None) | (Neg, Neg) | (Pos, Neg) |
| **Oct** | (Pos, Pos) | (None, None) | (Neg, Neg) | (Pos, Pos) |
| **Nov** | (Neg, Pos) | (None, None) | (Pos*, Neg) | (Pos, Pos) |
| **Dec** | (Pos*, Pos*) | (Pos*, None) | (Neg, Neg) | (Pos, Pos*) |

*Significant with a 95% confidence level. For logistic regression,
the positive/negative results are calculated using the average coefficient as the benchmark.*

Monthly statistical analysis is the simplest method. According to this method first I calculated the mean monthly returns for each month. Next, I divided the mean values by the standard deviations of the means to see which months offer abnormally positive or negative returns. The results for BIST 30 data suggest April and July support abnormally positive months. May and July offer abnormally negative returns for BIST 30. The results for S&P 500 data suggest March and December are significant outperformers.

The multiplicative decomposition analysis decomposes the data into trend and seasonal components. The trend equations for both BIST 30 and S&P 500 nominal data have positive coefficient on the time component. This is in line with the historical behavior of the stock indices as they grow exponentially over time. The decomposition analysis tends to overestimate the monthly effects for the first half of the year and underestimate these effects for the second half of the year. This is likely due to the positive trend component. November is the only month which supports a distinguishingly positive return for the BIST 30 data. No seasonality is observed in the S&P 500 data.

The dummy regression for BIST 30 is estimated using only April, May, August, and December binaries for BIST 30 data and April binary for S&P 500 data. For BIST 30 data April and December are expected to have strikingly positive returns, whereas the model suggests May and August have negative returns. For S&P 500, April supports a positive return. As we applied a stepwise regression procedure to include only the most relevant variables, all monthly effects are statistically significant.

In the logistic regression we estimated the odds for experiencing a positive return. December seems to be a great month for both BIST 30 and S&P 500 as the odds of earning a positive return are high for both markets. For BIST 30 the logistic regression is in line with the monthly statistical analysis of data as well as dummy variable regression. May and August support negative returns. For S&P 500, while all months support a lower odds ratio compared to December, July is the only month with a significantly negative return compared to December.

## 6. CONCLUSIONS

Can we guess what is next for the stock markets? Fama's (1970) efficient market theory suggests that in any efficient market it is impossible to make abnormal returns, but investors, speculators, arbitrageurs, and even academicians are looking for ways to outperform the crowds. In this article, I looked for one possible edge to see whether investing in some months are better than others. Four different methods are utilized, some of which offered similar results. For the BIST 30 data, one can claim a positive return for the months of April, July, and December, and negative returns for the months of May and August. For the S&P 500 data the only market anomalies are the positive returns on March and December.

In the aftermath of the financial crises, the role of extreme observations needs to be more emphasized. When one is looking for average returns over monthly periods, just a single extreme observation can mislead the results. Therefore, classifying the returns into binary variables (1 for positive, 0 for negative) and then analyzing the factors that result in positive events can give more robust results. I have done this binarization in the logistic regression. Of course, it is also possible and probably will give better estimates, if we extend the data both geographically and also over time. As the financial markets are highly integrated, inclusion of more indices around the globe will greatly benefit future research.

This research may or may not apply to other markets. Different research that utilize different time periods and applied in distinct stock markets can give divergent results. While global markets are highly interconnected with each other, each market has its own market-specific conditions. Also, in technical analysis, we are making a critical assumption that the past data is representative of future. This assumption is a very strong one in a highly volatile market where structural changes might happen over time. Adaptive market hypothesis offers some sort of explanation to the calendar anomalies (Urquhart and McGroarty, 2014), but the theoretical foundation of those persisting calendar effects is still not clear.

Behavioural finance can explain the persistent calendar anomalies to some extent. In order for the market anomalies to disappear, we need to have an efficient market where agents behave rationally. However, as Daniel and Titman (1999) explain the market actors have a tendency to behave irrationally during some periods. This irrational behaviour is also supported by the behavioural finance experiment conducted by Braga et al. (2009). The authors point that previous experiences drive the prices for the future bids. In the stock market this result implies that investors are likely to sell their stocks if they experienced losses during the past cycle and buy them if they experienced positive returns during another period. This kind of behaviour might explain the persistence of calendar effects.

In any case, if the past movements of the markets are completely random, then it is not logical trying to develop models that claim outperformance. However, if stock market indices have a memory which tends to repeat itself over time, then it might be possible to outperform the market. The idea where past behaviour of data can be utilized to forecast future movements is sometimes referred to as technical analysis. According to technical analysts, past behaviour of data is likely to repeat itself if the conditions that determine this behaviour happen again. However, it is not certain whether the same conditions do repeat themselves over the same period. Perhaps the behaviour of the markets is best explained with Mark Twain's (1894) famous quote in Pudd'nhead Wilson where he suggests every month is a dangerous month to invest in stock markets.